# Non-perturbative renormalization of the $\Delta S = 2$ four-fermion operator and the heavy-light static axial current.

A. Donini[a b], V. Giménez[c], G. Martinelli[a b], C.T. Sachrajda[d], M. Talevi[a] * and A. Vladikas[e]

[a]Dip. di Fisica, Universitá di Roma "La Sapienza" and I.N.F.N., Sezione di Roma I,
Piazzale Aldo Moro 2, I-00185 Roma, Italy.

[b]Theory Division, CERN, 1211 Geneva 23, Switzerland.

[c]Dep. of Theoretical Physics and IFIC, Universitat de València, E-46100 Burjassot (Valencia), Spain.

[d]Dep. of Physics, University of Southampton, Southampton SO17 1BJ, U.K.

[e]Dip. di Fisica, Universitá di Roma "Tor Vergata" and I.N.F.N., Sezione di Roma II,
Via della Ricerca Scientifica 1, I-00133 Roma, Italy.

We apply a recently introduced non-perturbative renormalization method to two types of lattice operators: the $\Delta S = 2$ four fermion operator and the heavy-light static axial current, which are relevant for the physics of $K$ and $B$ mesons respectively. The results of the non-perturbative calculations of the renormalization constants are compared with the corresponding perturbative ones.

## 1. Introduction

We present the results of two applications of a general method for non-perturbative (NP) renormalization of generic composite operators [1] (see also [2]). The method completely avoids the use of lattice perturbation theory (PT), known to be a source of uncertainty in the calculation of hadronic matrix elements [3].

We apply the general method to the $\Delta S = 2$ four-fermion operator $O^{\Delta S=2}$, relevant for $B_K$, and to the heavy-light static axial current $A_\mu$, relevant for $f_B$. Both $B_K$ and $f_B$ are very important for the study of CP violation. Moreover, the $\Delta S = 2$ operator is important as a preliminary study of the operators involved in the $\Delta I = 1/2$ rule. Indeed, while in the $\Delta S = 2$ case the chiral symmetry-breaking Wilson term induces mixing only with equal dimension operators, computable in PT, the $\Delta I = 1/2$ case presents mixing with lower dimension operators, implying power-divergent NP subtractions.

The NP method introduced in ref. [1] completely avoids the use of lattice PT, but not that of continuum PT, on which Wilson's OPE is based. The NP renormalization condition is applied directly to the Green functions of quarks and gluons. Given a bare operator $O^{\text{latt}}(a)$, the renormalized operator obtained with the NP method,

$$O_{\text{RI}}(\mu) = Z^{\text{latt}}_{\text{RI}}(\mu a) O^{\text{latt}}(a), \quad (1)$$

depends on the external states and the gauge, but not on method used to regulate the ultra-violet divergences. To stress this point, we call the NP renormalization scheme Regularization Independent (RI) [4]. The physical operator

$$O^{\text{phys}}(M) = C_{\text{RI}}(M/\mu) O_{\text{RI}}(\mu) \quad (2)$$

is independent of external momenta and gauge (up to higher orders in PT and lattice systematic effects) if the Wilson coefficient function $C_{\text{RI}}(M/\mu)$ in the RI scheme is calculated with the same external momenta and gauge of (1). The NP method is valid for any composite operator, as long as we can can find a window in the range of $\mu$ such that $\Lambda_{\text{QCD}} \ll \mu \ll O(1/a)$ [1].

To gain further insight on the size of the higher-order corrections, we will compare the NP renormalization constants with one-loop PT, using as our expansion parameter both the standard bare coupling $\alpha_s^{\text{latt}} = g_0^2(a)/4\pi$ (SPT) and a

---

*Talk presented by M. Talevi

| $\mu^2 a^2$ | $Z_+$ | $Z_1$ | $Z_2$ | $Z_3$ |
|---|---|---|---|---|
| 0.46 | 0.91(5) | 0.08(14) | 0.34(3) | 0.34(7) |
| 0.96 | 0.84(3) | 0.14(7) | 0.30(2) | 0.24(4) |
| 2.47 | 0.85(2) | 0.22(6) | 0.33(2) | 0.23(3) |
| SPT | 0.91 | 0.12 | 0.12 | 0.12 |
| BPT | 0.84 | 0.21 | 0.21 | 0.21 |

Table 1
Values of $Z_+, Z_{1,2,3}$ for several renormalization scales $\mu^2 a^2$. The PT values are at $\mu^2 a^2 = 1$.

"boosted" $\alpha_s^V \simeq 1.68\,\alpha_s^{\text{latt}}$, at $\beta = 6.0$ (BPT) [3].

## 2. $\Delta S = 2$ operator

We consider the renormalization of the four-fermion operator

$$O^{\Delta S=2} = (\bar{s}\gamma_\mu^L d)(\bar{s}\gamma_\mu^L d)\,,\ \gamma_\mu^L = \frac{1}{2}\gamma_\mu(1-\gamma_5)\,, \quad (3)$$

which appears in the weak effective Hamiltonian relevant for $K^0$-$\bar{K}^0$ mixing.

The discretization of the quark action à la Wilson, induces a mixing of the operator (3) with dimension-six operators of different chirality, $O_{SP}, O_{VA}, O_{SPT}$, determined by CPS symmetry (see refs. [5, 6] for their expressions). We define the renormalized operator as

$$O^{\Delta S=2}_{\text{RI}}(\mu) = Z_+(\mu a) O^{\Delta S=2}_{\text{sub}}(\mu a) \quad (4)$$
$$\equiv Z_+(O^{\Delta S=2} + Z_1 O_{SP} + Z_2 O_{VA} + Z_3 O_{SPT})\,.$$

Since there are no $\Delta S = 2$ operators of dimension lower than six, the mixing constants $Z_{1,2,3}$ are finite, while the logarithmically divergent $Z_+$ renormalizes multiplicatively the subtracted operator $O^{\Delta S=2}_{\text{sub}}$. The Wilson coefficient function in the RI scheme relating the operator (5) to the physical operator, necessary for a NP determination of $B_K$, can be found in ref. [8].

We determine the mixing constants $Z_i$ with a projection method on the four-point amputated Green functions [6]. Denoting by $O_i$, $i = 0, \ldots, 3$ the operators $O^{\Delta S=2}, O_{SP}, O_{VA}, O_{SPT}$, we define the projectors $\hat{P}_i$ $i = 0, \ldots, 3$ on the amputated tree-level Green functions $\Lambda_i^{(0)}$ of the operators $O_i$, i.e. $\text{Tr}\,\hat{P}_i \Lambda_j^{(0)} = \delta_{ij}$. The mixing constants $Z_i$ are fixed by the condition that the subtracted operator $O^{\Delta S=2}_{\text{sub}}$ is proportional to the bare free operator, i.e.

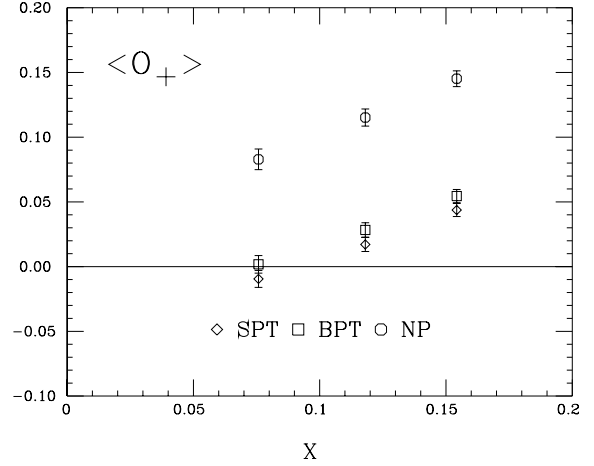

Figure 1. Chiral behaviour of $\langle O_+ \rangle$ as a function of $X$ (see text). The NP $Z$'s are at $\mu^2 a^2 = 0.96$.

$$\text{Tr}\,\hat{P}_k \Lambda^{\Delta S=2}_{\text{sub}}(pa) = 0,\ k = 1,2,3, \quad (5)$$

where $\Lambda^{\Delta S=2}_{\text{sub}}(pa)$ is the amputated Green function of $O^{\Delta S=2}_{\text{sub}}$, which is calculated at equal external momenta $p$ and in the Landau gauge in a completely NP fashion from numerical simulations. Eq. (5) yields a linear non-homogeneous system, from which we determine the mixing constants $Z_i$. The overall renormalization constant $Z_+$ is then determined by [6]

$$Z_+(\mu a) Z_q^{-2}(\mu a) \Gamma^{\Delta S=2}_{\text{sub}}(pa)|_{p^2=\mu^2} = 1, \quad (6)$$

where $\Gamma^{\Delta S=2}_{\text{sub}}(pa) = \text{Tr}\,\hat{P}_0 \Lambda^{\Delta S=2}_{\text{sub}}(pa)$, and $Z_q$ is the light-quark renormalization constant, determined NP from the conserved vector current [1].

We have done a feasibility study of the NP method on an ensemble of 36 configurations, on a $16^3 \times 32$ lattice, at $\beta = 6.0$, and with a single value of hopping parameter $\kappa = 0.1425$ for the SW-Clover quark propagator, in the lattice Landau gauge. We have performed the calculation at different scales $\mu^2 a^2$, spreading from 0.46 to 2.47. In tab. 1, we show the NP values of the $Z$'s for several representative values of $\mu^2 a^2$, and compare them with the PT result in the same gauge and external momenta [6]. Even with low statistics, for $\mu^2 a^2 \gtrsim 1$ the $Z$'s have relatively small errors. As expected the higher-order contributions differentiate the mixing constants with respect to the PT value, which at one-loop is the same for all three [5, 7]. We note that $Z_2$ is less scale-dependent than $Z_1$ and $Z_3$, but higher statitics is necessary to determine if the variations with $\mu^2 a^2$ are real or due to statistical fluctuations.

| $\mu^2 a^2$ | $\alpha$ | $\beta$ | $\gamma$ |
|---|---|---|---|
| 0.46 | 0.022(16) | 0.23(19) | 0.78(13) |
| 0.96 | 0.017(13) | 0.21(17) | 0.70(12) |
| 2.47 | 0.022(14) | 0.23(17) | 0.72(12) |
| SPT | −0.067(12) | 0.17(15) | 0.62(11) |
| BPT | −0.054(12) | 0.17(15) | 0.62(11) |

Table 2
Values of $\alpha, \beta$ and $\gamma$ for several renormalization scales $\mu^2 a^2$. The PT values are at $\mu^2 a^2 = 1$.

In order to investigate the effects of the NP corrections, we have combined our results with the high-statistics computation of the lattice matrix elements of the four-fermion operators $O_i$, $i = 0, \ldots, 3$, done in ref. [8]. In fig. 1, we show the chiral behaviour of $\langle O_+ \rangle = \langle \bar{K}^0 | O^{\Delta S=2} | K^0 \rangle_{\text{latt}} / \langle P_5 \rangle$ with the meson at rest as a function of $X = 8/3 f_K^2 m_K^2 / \langle P_5 \rangle^2$, where $\langle P_5 \rangle = |\langle 0 | \bar{s} \gamma_5 d | K^0 \rangle|$. Parametrizing the lattice matrix element as

$$\langle \bar{K}^0 | O^{\Delta S=2} | K^0 \rangle = \alpha + \beta m_K^2 + \gamma (p \cdot q) + \ldots \quad (7)$$

we found that the use of the NP $Z$'s leaves the values of $\beta$ and $\gamma$ almost unchanged compared to those obtained by using the one-loop PT values (both the PT and NP values of $\beta$ are compatible with zero). In contrast, $\alpha$ changes sign and its absolute value is reduced by about a factor of three in the NP case, becoming compatible with zero. This happens for any choice of $\mu^2 a^2$ between 0.46 and 2.47. The values of $\alpha, \beta$ and $\gamma$ for some representative scales of $\mu^2 a^2$ are shown in tab. 2. At present, this is the best test of the stability of the results which can be done, because the $Z$'s and the matrix elements have been computed on dif-ferent sets of configurations. Since $\alpha$ should vanish in the continuum limit, we conclude that the use of the NP $Z$'s improves the chiral behaviour for a large range of values of $\mu^2 a^2$ [6, 8].

## 3. Heavy-light axial current

Consider now the lattice heavy-light axial current in the static limit

$$A_\nu = \bar{b} \gamma_\nu \gamma_5 q, \quad (8)$$

where $b$ is a static heavy quark and $q$ is a propagating light quark. The NP renormalization of (8) is important in improving the accuracy in the determination of $f_B$ in the static limit, and consequently also at the physical value of $m_b$.

We renormalize $A_\nu$ by imposing [9]

$$Z_A^{\text{latt}}(\mu_0 a, \mu a) Z_b^{-1/2}(\mu_0 a) Z_q^{-1/2}(\mu a)$$
$$\Gamma_A(pa, p_0 a)|_{p_0 = \mu_0, p^2 = \mu^2} = 1, \quad (9)$$

where $Z_b$ and $Z_q$ are the heavy- and light-quark field renormalization constants respectively. $Z_b$ is defined NP from a conserved scalar current. In contrast to the light-light axial current, in the static case $A_\nu$ is not finite, so we have no Ward Identity to impose [1]. Moreover, we have two renormalization scales, $\mu$ and $\mu_0$, instead of one, due to the fact that Eichten's static formulation of the heavy quark is non-covariant. We will fix one of the two scales, e.g. $\mu_0$, and plot $Z_A$ as a function of the other. The physical renormalization constant $Z_A$ is

$$Z_A(m_b a) = C_{\text{RI}}(m_b/\mu_0, m_b/\mu) Z_A^{\text{latt}}(\mu_0 a, \mu a), (10)$$

where $C_{\text{RI}}$ must be computed, as $Z_A^{\text{latt}}$, in the RI scheme in continuum PT. The calculation of $C_{\text{RI}}$ at the next-to-leading order, necessary for a NP evaluation of $f_B$, can be found in [9].

In fig. 2 we show $Z_A^{\text{latt}}$ obtained with the NP method as a function of $\mu^2 a^2 = p^2 a^2$, with $\mu_0 a = p_0 a = 1, 2$. They have been calculated on an ensemble of 200 configurations, on a $18^3 \times 32$ lattice, at $\beta = 6.0$. The SW-Clover quark propagator has been computed in the lattice Landau gauge, at 3 values of the hopping parameter, $\kappa = 0.1425, 0.1432, 0.1440$, and extrapolated to the chiral limit $\kappa_{cr} = 0.1455$.

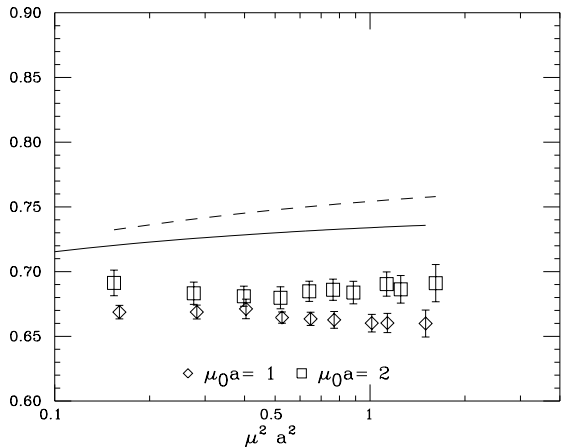

Figure 2. $Z_A^{\text{latt}}$ as a function of $\mu^2 a^2$ with $\mu_0 a = 1, 2$. The solid line is from BPT at $\mu_0 a = 1$, while the dashed line from BPT at $\mu_0 a = 2$.

In order to improve the off-shell matrix elements, we have also used an improved $O(a)$ heavy-quark propagator, which is simply obtained by multiplying the unimproved propagator in configuration space by [10, 9]

$$f(t) = 1 - \left(\frac{1}{3}\right)^{t+1}. \tag{11}$$

Moreover, to obtain a better quality of the signal for the Green functions, we have made use of traslational invariance and inserted the axial current in an arbitrary point, averaging over its positions. This method involves the evaluation of the heavy-quark propagator between two arbitrary times, and the computational effort required is reduced by the use of factorization properties of the heavy-quark propagator [9]. Although in the limit of infinite number of configurations, this method is equivalent to the usual method of inserting the current in the origin, we found a much better signal. The stability of the signal is shown in fig. 2, in which we also report the BPT result, calculated in the Landau gauge and with the same external momenta of the NP result. We note that the NP result lies somewhat below the BPT one, over the entire scale range. We would have an even greater discrepancy with SPT. This is not surprising since the one-loop corrections are quite large [11].

## 4. Conclusions

The results of the NP renormalization applied to logarithmically divergent operators, although preliminary, are encouraging. In particular, the chiral behaviour of the $\Delta S = 2$ operator seems to be significantly improved. This motivate us to repeat the calculation, with larger statistics for the $\Delta S = 2$ operator and with unimproved Wilson fermions for the axial current. We plan to apply the NP method to power-divergent operators, relevant for the $\Delta I = 1/2$ rule.